# Switching Graphitic Polytypes in Elastically Coupled Islands

Nirmal Roy[1], Simon Salleh Atri[1], Yoav Sharaby[1], Noam Raab[1], Youngki Yeo[1], Watanabe Kenji[2], Takashi Taniguchi[3], Moshe Ben Shalom[1]

[1]School of Physics and Astronomy, Tel Aviv University, Tel Aviv, Israel
[2]Research Center for Functional Materials, National Institute for Materials Science, Tsukuba, Japan
[3]International Center for Materials Nanoarchitectonics, National Institute for Materials Science, Tsukuba, Japan

Van der Waals polytypes are commensurate configurations of two-dimensional layers with discrete crystalline symmetries and distinct stacking-dependent properties[1]. In graphitic polytypes, the different stacking arrangements of graphene sheets exhibit rich electronic phases, such as intrinsic electric polarizations[2], orbital magnetizations[3,4], superconductivity[5,6], and anomalous fractional Hall states[7,8]. Switching between these metastable periodic configurations by controlling interlayer shifts unlocks intriguing multiferroic responses[9].

Here, we report super-lubricant arrays of polytypes (SLAP) devices[10], with nanometer-scale islands of Bernal polytypes that switch into Rhombohedral crystals and vice versa under a shear force as low as 6 nano-Newtons. We assemble these four-layer SLAP structures by aligning a pair of graphene bilayers above and under circular cavities in a misaligned spacer layer. Using local current measurements, we detect the shifts between the active bilayers and reveal long-range elastic relaxations outside the cavities that enable efficient nucleation and spontaneous sliding of stacking dislocation inside the islands. We demonstrate configurable, deterministic, and robust polytype switching by confining these boundary strips in narrow cavity channels that connect the islands. Such controlled switching between elastically-coupled single-crystalline islands is appealing for novel multiferroic SlideTronic applications.

The quest to transform a given crystalline material into another with desirable properties has proven challenging throughout history. For example, centuries-long attempts by alchemists to transmute copper into gold were doomed to fail, unaware that such a transformation requires a nuclear reaction. In contrast, graphite and diamond crystals comprise the same carbon atoms, simply rearranged into different structures. Still, transforming graphite into diamond requires extreme temperatures and pressures to break and rearrange the tight covalent bonds, hence remaining impractical[11,12]. A more attainable objective is switching between distinct polytype configurations within layered crystals like graphite by shifting the carbon planes between relatively weak van der Waals (vdW) stacking arrangements[13,14]. These room-temperature sliding transitions were recently controlled using feasible electric fields in a broad family of polar vdW polytypes made of non-polar layers of binary compound[15,16] such as transition metal dichalcogenides (TMDs)[17] and boron nitride (BN)[18–20]. The observed interfacial ferroelectricity and cumulative polarizations, co-existing with planar conductivity[21,22] are appealing to modern technologies. The latter, however, require a rapid, efficient, and local control of the atomic shifts, as in electronic transistors. From this perspective, switching between graphitic polytypes is potentially very profitable[9,16,23], despite having relatively small changes in the structural-dependent electronic bands[24–27] (compared with graphite to diamond transition). The main challenge is maintaining robust structural (meta)stabilities while facilitating efficient and rapid sliding instabilities for smooth switching.

To this end, it is vital to understand the microscopic mechanism behind these discrete interlayer shifts. A first insight into these questions was recently obtained by monitoring the internal electric polarization of BN domains before and after the switching[19]. It was shown that external electric fields push preexisting boundary stripes between adjacent structural configurations rather than causing a rigid layer motion[19,28–31]. By sliding the boundary strips in the plane, the vertical field extends stacking domains with co-aligned polarization over stacking domains with opposite orientations. Consequently, the energy cost associated with nucleating, deforming, pinning, and sliding these partial dislocation strips determines the structural response[9]. Switching between uniform single-crystalline structures without preexisting domains and strips requires abrupt stimuli, such as large mechanical forces or intense electric currents, to nucleate a local boundary strip and then elongate it across the structure dimensions[32].

To overcome this boundary nucleation barrier, we have recently introduced a super-lubricant array of polytypes (SLAP) device concept[10], demonstrating purely electrical nucleation of boundary strips and efficient control of structural transitions in uniform, single-crystalline nm-scale islands. The idea is to reduce the boundary strips' energy cost by confining the commensurate island diameter without physically cutting the active layers. Instead, we design the islands' shapes and dimensions by etching cavities into incommensurate spacers and allowing the active layers to touch and form commensurate polytypes only within the cavities. This cavity approach eliminates the formation of structural pining barriers by edge-dangling atoms that tend to zip the active layers together and prevent interlayer motions (see Fig.1). Furthermore, the minute friction with the incommensurate twisted spacer outside the cavity[33] supports long-range (and low energy cost) elastic relaxations of the interlayer shifts that accompany switching events. This structural lubricity[34] at the misaligned medium between the islands enables elastic inter-island interactions and coupled switching responses, unavailable in standard information technologies based on 3D crystals. Ahead, we report electrical currents measurements with a nanometer-scale resolution of conductive graphite SLAP devices that exhibit robust stability of single-crystalline Rhombohedral and Bernal polytypes and switching energies as low as $10^{-15}$ Joules in islands as small as 30 nm.

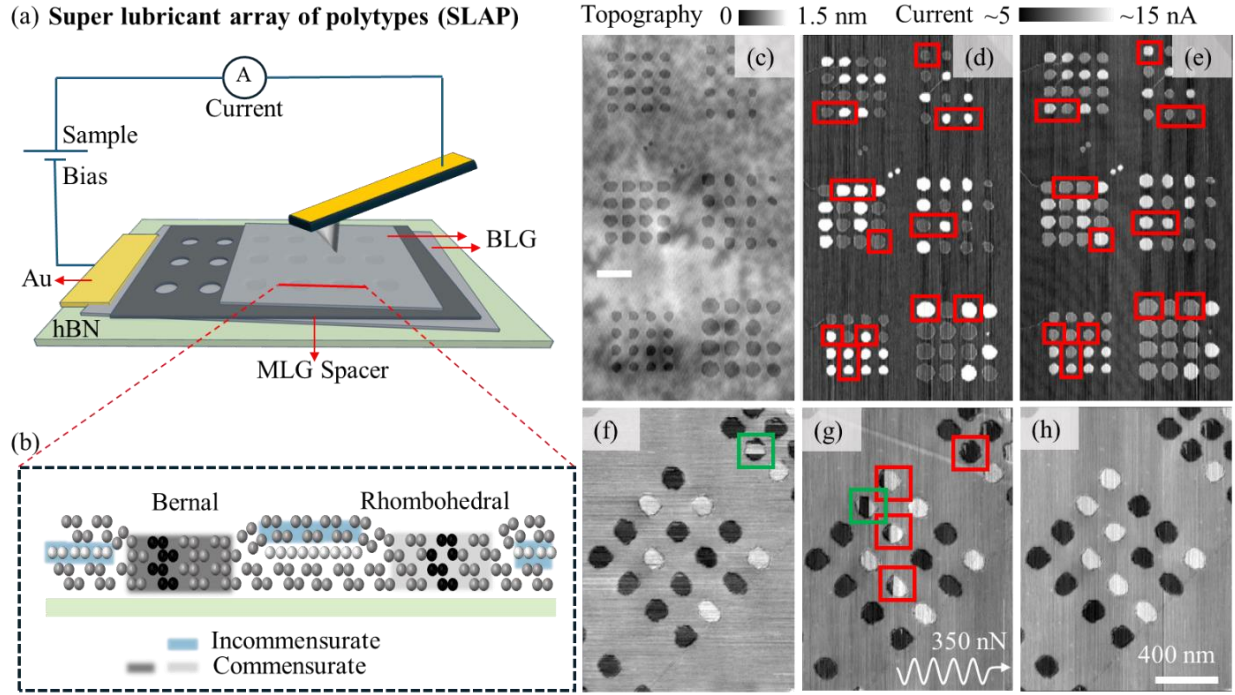

*Figure 1: Switchable graphitic polytypes in superlubricant arrays.* **a** *Illustration of a SLAP device and the conductive AFM measurement circuit. A rotationally misaligned monolayer graphene spacer (dark gray) with circular cavities is sandwiched between a pair of parallel bilayer graphene sheets (light gray), placed on a flat hexagonal boron-nitride substrate.* **b** *Cross-sectional view of two cavities along the red line-cut in (a). Black commensurate spheres within each cavity (shaded in dark and light gray) mark an eight-atom unit cell of Bernal (B) and Rhombohedral (R) polytypes. Lubricant incommensurate interfaces outside the cavities are shaded in light blue.* **c** *AFM topography image.* **d**. *Imaging current scan measured with low loading force (100 nN) on the tip. Dark and bright islands are B and R polytypes, respectively.* **e**. *Imaging scan after performing several high-load switching scans. Red and green frames mark single and double-switching events, respectively.* **f-h** *Example of another SLAP array. The current map during the switching scan with a 350 nN load is shown in (g). Scale bars mark 400 nm.*

**Imaging Switchable SLAP**

As a model graphitic SLAP system, we use a pair of aligned bilayer graphene to encapsulate a twisted monolayer spacer (see Fig.1a, Fig.S1, and SI.1). The spacer layer is patterned with circular cavities with 30 to 500 nm diameters arranged in rectangular arrays with 30 to 1000 nm separations between the cavities. The cross-sectional illustration in Fig.1b shows cavities with commensurate tetra-layers of Bernal (B) and rhombohedral (R) polytypes. Conversely, the bilayers outside the cavities are incommensurate and face a misaligned super lubricant surface (Fig.1b). Fig.1c shows an atomic force microscope (AFM) topography map of a typical sample with dark circles corresponding to a height drop of 0.33 nm, the thickness of the monolayer spacer.

Current maps between the AFM tip and a metallic gold (Au) lead connected to the active layers are presented in panels d-h (see the electric circuit illustration in panel a). These current imaging maps are measured at a typical 5 mV sample bias (with ~5 MΩ circuit resistance) and a vertical load force of 100 nN. Notably, all observed cavities divide into bright islands representing higher currents, corresponding to Rhombohedral (R) polytype, and dark islands with ~5 nA suppression of the current, corresponding to Bernal (B) polytypes. We confirm this correspondence by measuring ~ 500 cavities in 20 arrays assembled on five samples and comparing the current signals with our previous Raman and surface potential measurements[2,35] (see Fig.S3 and SI.4). We note that the precise circuit resistance fluctuates during the scan due to occasional attachments of surface contaminants to the tip apex. Nevertheless, islands of the R polytype remain brighter than B islands, allowing us to determine the local configuration. Fig.1 panels d and e show imaging scans with relatively low ~100 nN vertical load forces taken before and after a switching scan of the surface with a higher tip load of 300 nN. This load value correspond to a vertical pressure of 240 MPa, and a lateral friction force of 15 nN, as measured from the tips' torque displacements (see Fig.S2). Remarkably, ~ 15% of the islands switch from R to B or vice versa (marked in red frames) without any apparent boundary dislocations within the islands.

To monitor these switching dynamics, we measured the current before, during, and after the high-load scans. We observe instantaneous switching events at specific tip positions during the scan. Apparently, the boundary dislocation nucleates and then propagates spontaneously to switch the island entirely. For example, the red frames in the high-load scan, Fig.1g, show sharp color switching within the islands (during a single line scan), while uniform islands appear in the previous and following low-load imaging scans (panels f, h). The same response is observed if we stop the scan immediately after detecting a color jump and then take a low-load imaging scan. Occasionally, we observed two successive switching events within the same cavity (see green frames), resulting in the same polytype at the following imaging scan. Such spontaneous propagation and annihilation testify to a low energy barrier for boundary strip sliding, as enabled by the SLAP concept.

In contrast, samples containing aligned spacers show restrained strip dynamics due to stiff dislocation networks that form between the functional layers and the commensurate spacer (Fig.S4). The transition from low-barrier spontaneous strip motion to restricted domain patterns appears for spacers twisted by a few degrees. For a twist of 1.4 degrees we observe strips within the island which are dragged during the scan and slides back to their initial position once the dragging force is released (see panels c-g of Fig.S4). For smaller twist angles, the overall domain patterns and switching dynamics are rather complicated and depend on the island dimensions and misalignments between the active layers (see panels h-i of Fig.S4). Here, the planar reconstructions inside and outside the cavities enable a unique insight into the relative stability of the strips (see section SI.5).

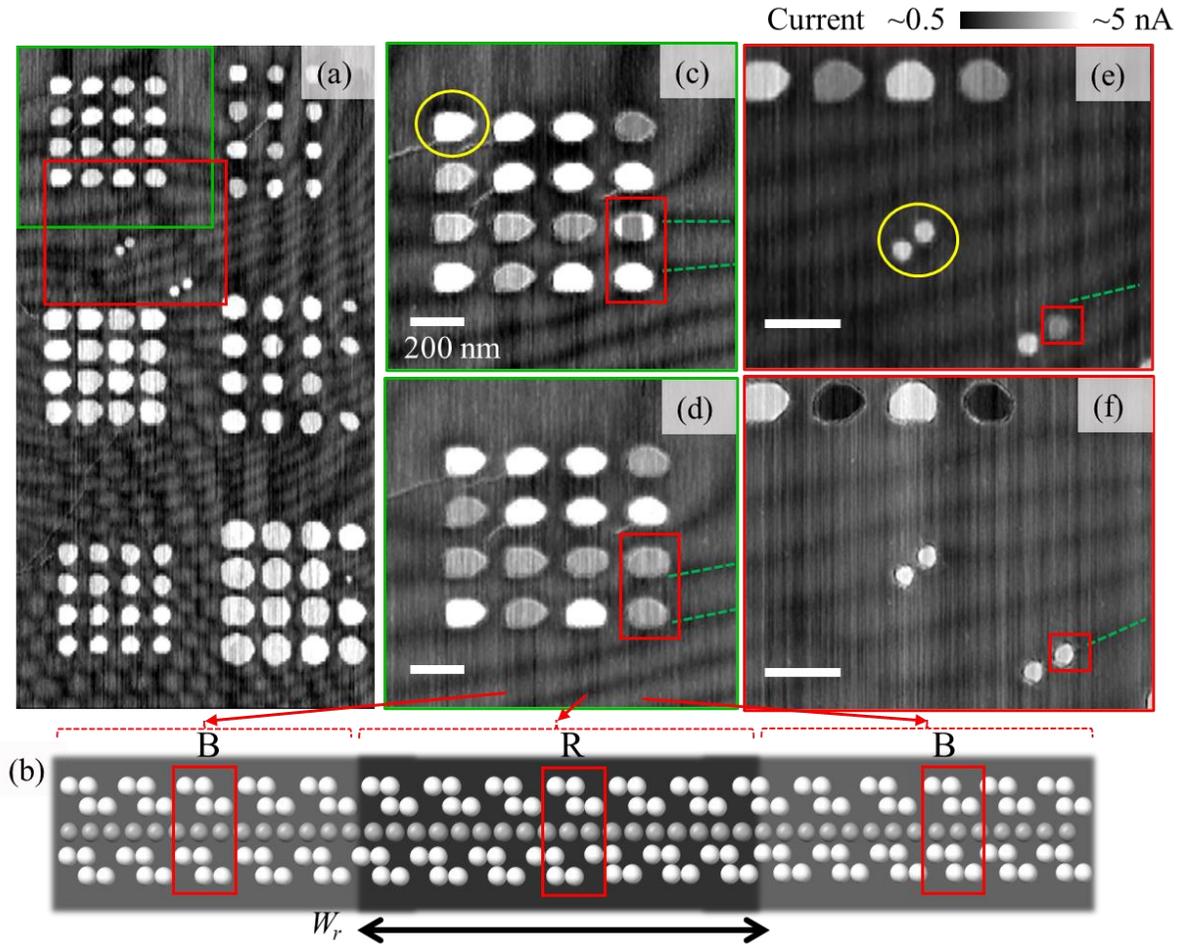

*Figure 2: Remote stacking boundary strips. **a** Current AFM map as in Fig.1d, optimized to detect partial dislocation between the separated active layers (dark stripes) outside the cavities. **b** Cross-sectional illustration of the remote stacking at the boundary strip. The unit cells of weakly commensurate Bernal-like (B) and Rhombohedral-like (R) remote configurations across the misaligned spacer are framed in red. **c, d** Zoom-in maps taken before and after R to B switching showing the accompanied motions of remote dislocation. The dashed green lines intersect the bright R islands and then shift to the bottom of the cavity while the islands switch to B polytypes. The yellow circle show an R island, which is not intersected by any remote dislocation. Instead, a dark circular halo appears around the island. **e, f** Example of remote dislocations which are attracted to the cavity and tend to align with the inter-island axis. The red frame shows a switching event of one individual island separated by 50 nm from another island.*

**Interlayer Shifts across the Spacer**

To monitor the interlayer shifts outside the cavities that accompany each switching event within the islands, we focus on SLAP devices with thin monolayer graphene spacers. Notably, current maps with a relatively high tip load of 400 nN reveal moiré patterns between the top and bottom bilayers across the misaligned spacer, see Fig.2a. These "remote" moiré patterns are formed due to planar structural relaxations between the active bilayers despite the incommensurate spacer, see the cross-sectional stacking illustration in panel b. Apparently, the residual interactions between the aligned bilayers expand bright domains of enhanced current over dark domains that squeeze into relatively narrow boundary strips. These better-stable bright domains correspond to remote B stacking, while the dark strips correspond to remote R stacking and other intermediate configurations. To see how, we point to the hexagonal shape of this remote moiré pattern at the bottom of the map. The center of the hexagon reveals the better-stable remote-stacking configuration. It is somewhat surprising since structural domains in marginally twisted successive layers are typically hard to distinguish. Usually, they form triangular shapes of similar areas at opposite sides of each boundary, as in Fig.S4. Triangle patterns appear for minor energy differences between commensurate configurations, which is insufficient to push the essential and expensive elongation of the boundary strips[36]. Conversely, the hexagonal pattern in Fig.2a means that strip elongations in the case of remote commensurate layers have a low energy cost. Since the dark remote boundary strips typically intersect R islands, while the strips shift away from B islands, we conclude that B is the better remote configuration. Panels c-f show examples of intersecting and shifted strips (marked in dashed green lines) and a corresponding switching from R to B islands, as expected. This fascinating interplay of expensive strips between successive layers inside the islands that continue as relatively cheap remote dislocations outside the cavities determines the reach moiré patterns observed in Fig.2 and Fig.S4.

Markedly, the width of remote stacking boundaries is $W_r \sim 60 \pm 15$ nm, much wider than the $W_s \sim 7$ nm wide boundary strips of successive layer [37]. These $W$ values are set by the energy cost associated with nucleating and elongating a strip and the elastic stress relaxation lengths, which are dominant factors in the switching dynamics. Each $W$ is determined by the energy gain of the commensurate stacking potential, the planar stiffness of the layers, and the Burgers shift at the boundary between the structural phases[38]. On the one hand, narrow $W$ reduces the number of atoms at the intermediate misaligned stacking configuration and the corresponding adhesion energy loss by $\propto E_{int} W/a$, with $E_{int}$ the misalignment energy cost per atomic area, and $a = 0.142$ nm the bond length. On the other hand, the elastic energy loss due to planar strains within each layer grows as $\propto a^2/W$ (here $a$ is also the Burgers shift vector). The sum of these two terms is minimized for $W_s \approx \frac{a}{2}\sqrt{\frac{E_Y}{E_{int}(1-v^2)}} \approx 7\ nm$ [37–39] for successive layers with $E_{int} \sim 2 meV/atom$ and $W_r \sim 60$ nm for remote stacking boundaries. $E_Y$ and $v$ are Young's modulus and Poisson's ratio, respectively. Hence, from the ratio $\frac{W_r^2}{W_s^2} = \frac{E_{int}}{E_{remote\_int}} \approx 75$ we find $E_{remote\_int} \approx 0.015$ meV per atom, and an overall remote strip energy cost of about 100 meV per nm length, ten times smaller than for successive dislocations. This relatively low energy cost enables easier elongations, deformations, and sliding of remote dislocations and subsequent switching of the polytype islands. Thicker spacers that completely mute the residual coupling between the layers, result in even cheaper rearrangements at the lubricant medium between the cavities.

Finally, we note the dark ring ($\sim W_r$ wide) at each cavity's edge that appears even in isolated cavities (see the yellow frames in Fig.2c for example). These circular strips of remote incommensurate regions appear because the top bilayers stretch while sagging into the cavity. They demonstrate a long-range stress relaxation length that substantially exceeds $W_s$ and provides a novel elastic interaction mechanism between adjacent islands (unlike the few nm scale relaxations in the commensurate spacers between the cavities, see Fig.S4). To further reduce the overall elastic energy, the remote boundary strips tend to align with the cavity pattern and merge with the circular stretch at the cavities' edge (see Fig.2e, f).

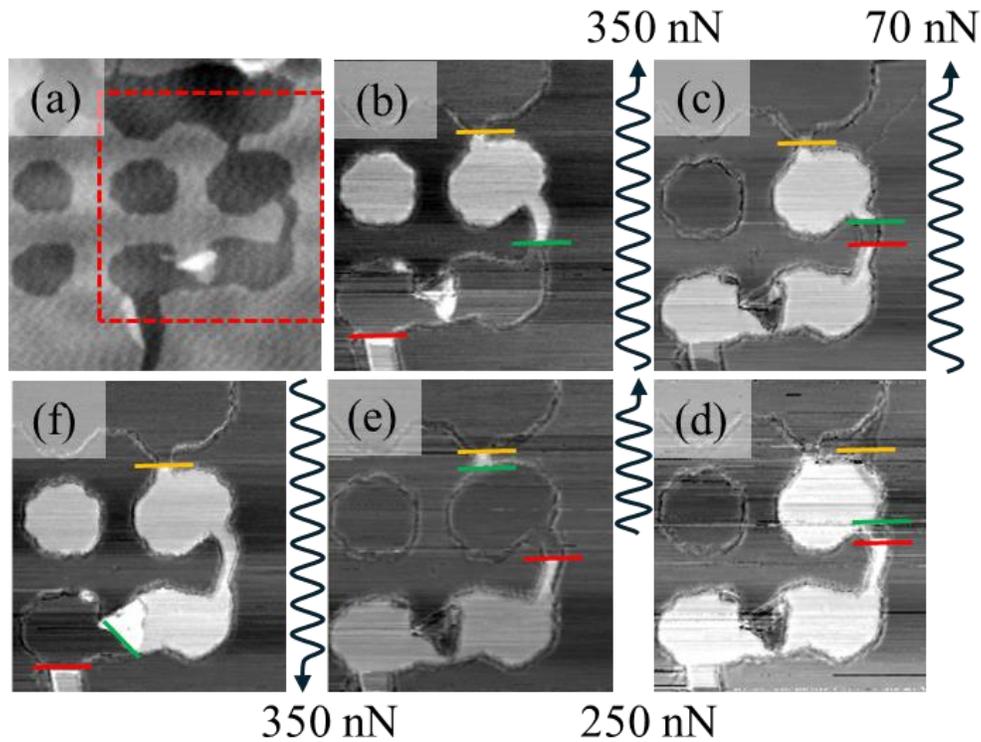

*Figure 3: Coupled cavities via narrow channels. a Topographic image of a lubricant cavity array connected by narrow cavity channels. b-f, Successive current imaging maps tracing the direct boundary strip motions (colored lines) before and after switching scans (not shown). The orientation and loading force of the switching scans between the low-load imaging scans are indicated by the black wiggling arrow. For Example, the low-load imaging scan in (b)is followed by a switching scan with a 350 nN load and a slow scan axis oriented from bottom to top, and then panel (c) show the final rearrangements.*

**Designing Inter-Island Coupling**

To lower the coercive friction forces required for switching, and enhance the coupled island response, we implement an additional design concept. Here, islands are linked by narrow channels that capture boundary dislocations and orient them perpendicular to the channel axis. These direct strips are then pushed to switch the island efficiently and deterministically (Fig.3). The narrow constrictions minimize the length and energy cost of the dislocation strips between successive layers while rearranging the less expensive remote dislocation pattern. Fig.3a shows a typical topography map of 20 nm wide channels that link 150 nm wide cavities (see the dark regions with no spacer). Panels b-f show current maps of the same region with boundary dislocations that appear only in the channels and always in transversal channel direction (marked in red, green, and orange lines). Although not directly detected under these imaging conditions, each direct boundary strip must proceed as a remote boundary into the lubricant medium.

To switch the polytype deterministically, we place the tip outside the nearby strip and scan the slow axis towards the cavity. Panel (c), for example, was imaged after placing the tip below the red boundary in panel (b) and pushing it up as sketched by the wiggling black line. For scanning loads exceeding ~ 300 nN, the direct (and remote) dislocations slide to cross the two bottom islands and switch the B polytype to R (from dark to bright cavities). We note that pushing dislocations along the channels without crossing the island requires much lower loading forces below 70 nN (corresponding to a 6 nN friction force only), as shown in panel (d), which is taken after the pushing scan sketched in (c). The maps in panels (d) and (e) complete a demonstration of four individually controlled islands by pushing the successive boundaries in all directions. Within the channel, the dislocations can either attract each other as in panel (e) (orange and green lines) or stay apart (panel c, green and red), with each local narrowing of the channel serving as a stabilizing pinning potential and a sliding barrier. We emphasize that the full sliding dynamics include reorientation and motion of remote dislocations outside the islands. For Example, the circular island in panel (c), which is not connected by any channel, was also switched after the scan sketched in panel (b).

**Conclusions and Outlook**

Graphitic SLAP devices may exhibit diverse switching responses with a variety of tuning knobs to design individual islands' switching as well as coupled structural dynamics in the cavity array. The incommensurate lubricant medium between the cavities is crucial for robust, deterministic, and efficient switching of commensurate polytypes within the islands. Conversely, devices with commensurate spacers exhibit stiff networks of boundary strips that prevent switching or, alternatively, revert the domain landscape to its initial pattern once the external force is removed (see SI.5 and Fig.S4). Increasing the cavity diameter enhances the islands (meta)stability and coercive switching force. For example, switching our smallest 30 nm holes requires coercive friction forces as small as 10 nN, while holes with a 300 nm diameter typically require more than 15 nN.

Designing the cavities' shapes and interconnecting channels further decreases the coercive force for more rapid and efficient switching. Once the structural pining from edge-open bonds is sufficiently low, the boundary strips shift spontaneously into narrow constrictions. Adjusting the broadening of these interconnect constrictions, the cavities diameter, and separation distance enables sensitive control of the switching energy and the inter-island couplings. In principle, the minimum energy required for each switching event is set by the strip elongation length with a typical cost of 1eV/nm[40]. The latter value translates to a ~$5\times10^{-18}$ Joules for a 30 nm island diameter, which is highly competitive compared to present nonvolatile ferroelectric technologies[41]. Previously, however, the global stacking rearrangements that must accompany any local switching of typical vdW structures involved substantial energy costs and required 4 µN friction forces. [32]. The later force value corresponds to $10^{-14}$ Joules of work performed by the tip over 30 nm line. Notably, in the present experiments, elongating the remote boundaries at the lubricant interfaces outside the cavity of SLAP devices is much cheaper. For example, the work performed by the 6 nN friction force to move a direct dislocation over 30 nm (including the remote rearrangements) is about $2\times10^{-16}$ Joules only.

Moreover, the distinct long-range strain relaxation observed outside the cavities, more than 60 nm for a spacer thicker than a monolayer, compared to less than 10 nm in commensurate interfaces[37], enables elastically coupled inter-islands switching. These couplings could enable neuromorphic-like applications in which islands serve as two-state bits and information stored in the network between the bits is manipulated using mechanical and electrical controls. In particular, direct strips captured in the channels between the islands can substantially increase the coupled switching response. While the SLAP concept extends to all vdW polytypes with various stacking-dependent properties, the present graphitic devices demonstrated stable switching to rhombohedral islands and hence further enable multiferroic electronic correlations. We also foresee novel dot structures in which the quantum dot is defined by the cavity shape rather than local electric gates or the physical dimensions of the active layers.

**Methods: Sample Preparation and Device Details:**

Our device structure consists of (from top to bottom); a bi-layer graphene (BLG) / misaligned monolayer graphene (MLG) spacer with a cavity array / aligned BLG / 50 nm thick hBN/ 90 nm $SiO_2$ / 0.5 mm Si substrate, fabricated using a dry polymer stamping method (See Ref[1,2] of SI). A large BLG is initially etched into two parts using high-intensity pulsed laser light (1064 nm, WITEC alpha300 Apyron, Fig.S1.a, d). In a separate step, we etch a cavity array into a MLG spacer using electrode-free local anodic oxidation (LAO) lithography (See Ref[3] of SI) (Fig.S1.b, e).

We assembled the device using a Polyvinyl alcohol (PVA)-coated polydimethylsiloxane (PDMS) stamp to pick up the hBN, the first BLG, the MLG spacer, and the second BLG. To flip the structure and place the active BLG at the top, facing the tip of the AFM, we use a second PDMS stamp coated with PMMA (Polymethyl methacrylate). Then we bring the stacked layers from the PDMS-PVA stamp into contact with the PMMA stamp and inject water into the gap to dissolve the PVA coating, facilitating the transfer. Finally, we drop the entire stack onto a $Si/SiO_2$ substrate at 150°C. After fabrication, we design electrical contact pads (Cr/Au) using electron beam lithography.

**Conducting Atomic Force Microscopy (C-AFM) Measurements:**

The current maps presented in this work were captured using a Park NX10 Hivac in C-AFM mode under an $N_2$ atmosphere. A HQ:NSC35/Pt tip with a spring constant of 5.4 N/m and a resonant frequency of ~150 kHz was used for all measurements. Prior to current mapping, the sample surface was thoroughly cleaned by performing multiple scans in contact mode with a loading force of 350 nN. This process removes any residual contaminations and confirms that all cavity arrays are properly sagged. Following the tip cleaning, a fresh AFM tip is utilized for the current imaging. To avoid saturation of the tip current due to low resistance in several specific devices, we add a series resistor (1-10 MΩ) into the measurement circuit.

**Friction Force Measurements**

We record the lateral photodiode signals during both forward and backward scans, as illustrated in Fig.S2.a, b, to measure the friction force. Friction force induces torsional motion in the AFM tip (Fig.S2.b), which manifests as a difference between the forward and backward lateral photodiode signals (measured in volts), as shown in Fig.S2.c and d. We generate a friction force map by calculating this difference, which reflects the torsional motion of the AFM tip. To convert the potential shift into a friction force, we use a SiO$_2$ substrate (with a known friction coefficient of 0.08) and a known conversion factor (nN/V) for lateral force (see Ref[4] of SI). Fig.S2.e shows a plot of the friction force as a function of the loading force, as we deduce from dedicated friction force maps (see examples in Fig.S2 panels f-h).

**Acknowledgments**

We acknowledge Prof. Yoav Lahini for many discussions, Neta Shivek Ravid, Itzhak Yakov Malker, and Peter Yanovich for laboratory support, and Andrea Cerreta from Park Systems for AFM support. K. W. and T. T. acknowledge support from the JSPS KAKENHI (Grant Numbers 21H05233 and 23H02052) and World Premier International Research Center Initiative (WPI), MEXT, Japan. M.B.S. acknowledges funding by the European Research Council under the European Union's Horizon 2024 research and innovation program ("SlideTronics", consolidator grant agreement no. 101126257) and the Israel Science Foundation under grant nos. 319/22 and 3623/21. We further acknowledge the Centre for Nanoscience and Nanotechnology of Tel Aviv University.

**References:**

1. Guinier, B. A., BoKIJ, G. B., Jagodzinski, H. & Abrahams, S. C. *HAHN (Ex Officio*. *Academy of Sciences of the USSR, Staromonetny* vol. 40 (1984).

2. Atri, S. S. *et al.* Spontaneous Electric Polarization in Graphene Polytypes. *Advanced Physics Research* **3**, (2024).

3. Shi, Y. *et al.* Electronic phase separation in multilayer rhombohedral graphite. *Nature* **584**, 210–214 (2020).

4. Zhou, H. *et al.* Half- and quarter-metals in rhombohedral trilayer graphene. *Nature* **598**, 429–433 (2021).

5. Zhou, H., Xie, T., Taniguchi, T., Watanabe, K. & Young, A. F. Superconductivity in rhombohedral trilayer graphene. *Nature* **598**, 434–438 (2021).

6. Pantaleón, P. A. *et al.* Superconductivity and correlated phases in non-twisted bilayer and trilayer graphene. *Nature Reviews Physics* **5**, 304–315 (2023).

7. Lu, Z. *et al.* Fractional quantum anomalous Hall effect in multilayer graphene. *Nature* **626**, 759–764 (2024).

8. Choi, Y. *et al.* Electric field control of superconductivity and quantized anomalous Hall effects in rhombohedral tetralayer graphene. *http://arxiv.org/abs/2408.12584* (2024).

9. Vizner Stern, M., Salleh Atri, S. & Ben Shalom, M. Sliding van der Waals polytypes. *Nature Reviews Physics* (2024) doi:10.1038/s42254-024-00781-6.

10. Yeo, Y. *et al.* *Switchable Crystalline Islands in Super Lubricant Arrays*.

11. Fahy, S., Louie, S. G. & Cohen, M. L. *Pseudopotential Total-Energy Study of the Transition from Rhombohedral Graphite to Diamond*. PHYSICAL REVIEW vol. 8 (1986).

12. Xie, H., Yin, F., Yu, T., Wang, J. T. & Liang, C. Mechanism for direct graphite-to-diamond phase transition. *Sci Rep* **4**, (2014).

13. Yankowitz, M. *et al.* Electric field control of soliton motion and stacking in trilayer graphene. *Nat Mater* **13**, 786–789 (2014).

14. Li, H. *et al.* Global Control of Stacking-Order Phase Transition by Doping and Electric Field in Few-Layer Graphene. *Nano Lett* **20**, 3106–3112 (2020).

15. Li, L. & Wu, M. Binary Compound Bilayer and Multilayer with Vertical Polarizations: Two-Dimensional Ferroelectrics, Multiferroics, and Nanogenerators. *ACS Nano* **11**, 6382–6388 (2017).

16. Wu, M. & Li, J. Sliding ferroelectricity in 2D van der Waals materials: Related physics and future opportunities. *Proceedings of the National Academy of Sciences of the United States of America* vol. 118 Preprint at https://doi.org/10.1073/pnas.2115703118 (2021).

17. Fei, Z. *et al.* Ferroelectric switching of a two-dimensional metal. *Nature* **560**, 336–339 (2018).

18. Woods, C. R. *et al.* Charge-polarized interfacial superlattices in marginally twisted hexagonal boron nitride. *Nat Commun* **12**, (2021).

19. Vizner Stern, M. *et al.* *Interfacial Ferroelectricity by van Der Waals Sliding*. https://www.science.org.

20. Yasuda, K., Wang, X., Watanabe, K., Taniguchi, T. & Jarillo-Herrero, P. *Stacking-Engineered Ferroelectricity in Bilayer Boron Nitride*. https://www.science.org.

21. Deb, S. *et al.* Cumulative polarization in conductive interfacial ferroelectrics. *Nature* **612**, 465–469 (2022).

22. Cao, W. *et al.* Polarization Saturation in Multilayered Interfacial Ferroelectrics. *Advanced Materials* **36**, (2024).

23. Wang, C., You, L., Cobden, D. & Wang, J. Towards two-dimensional van der Waals ferroelectrics. *Nature Materials* vol. 22 542–552 Preprint at https://doi.org/10.1038/s41563-022-01422-y (2023).

24. Aoki, M. & Amawashi, H. Dependence of band structures on stacking and field in layered graphene. *Solid State Commun* **142**, 123–127 (2007).

25. Koshino, M. & McCann, E. Multilayer graphenes with mixed stacking structure: Interplay of Bernal and rhombohedral stacking. *Phys Rev B Condens Matter Mater Phys* **87**, (2013).

26. Bao, W. *et al.* Stacking-dependent band gap and quantum transport in trilayer graphene. *Nat Phys* **7**, 948–952 (2011).

27. Koshino, M. Stacking-dependent optical absorption in multilayer graphene. *New J Phys* **15**, (2013).

28. Ko, K. *et al*. Operando electron microscopy investigation of polar domain dynamics in twisted van der Waals homobilayers. *Nat Mater* **22**, 992–998 (2023).

29. Gao, Y. *et al*. Tunnel junctions based on interfacial two dimensional ferroelectrics. *Nat Commun* **15**, (2024).

30. Bian, R. *et al. Developing Fatigue-Resistant Ferroelectrics Using Interlayer Sliding Switching*. https://www.science.org.

31. Yasuda, K. *et al. Ultrafast High-Endurance Memory Based on Sliding Ferroelectrics*. https://www.science.org.

32. Jiang, L. *et al*. Manipulation of domain-wall solitons in bi- and trilayer graphene. *Nat Nanotechnol* **13**, 204–208 (2018).

33. Dienwiebel, M. *et al*. Superlubricity of graphite. *Phys Rev Lett* **92**, (2004).

34. Müser, M. H. Structural lubricity: Role of dimension and symmetry. *Europhys Lett* **66**, 97–103 (2004).

35. Nirmal Roy, et. , al. Graphitic Polytypes in Penta-layer Graphene. *(under Preparation)*.

36. Weston, A. *et al*. Atomic reconstruction in twisted bilayers of transition metal dichalcogenides. *Nat Nanotechnol* **15**, 592–597 (2020).

37. Alden, J. S. *et al*. Strain solitons and topological defects in bilayer graphene. *Proc Natl Acad Sci U S A* **110**, 11256–11260 (2013).

38. Kushima, A., Qian, X., Zhao, P., Zhang, S. & Li, J. Ripplocations in van der Waals layers. *Nano Lett* **15**, 1302–1308 (2015).

39. Popov, A. M., Lebedeva, I. V., Knizhnik, A. A., Lozovik, Y. E. & Potapkin, B. V. Commensurate-incommensurate phase transition in bilayer graphene. *Phys Rev B Condens Matter Mater Phys* **84**, (2011).

40. Lebedeva, I. V., Lebedev, A. V., Popov, A. M. & Knizhnik, A. A. Dislocations in stacking and commensurate-incommensurate phase transition in bilayer graphene and hexagonal boron nitride. *Phys Rev B* **93**, (2016).

41. *IEEE International Roadmap for Devices and Systems, "Beyond CMOS and Emerging Materials Integration." Institute of Electrical and Electronics Engineers, 2023, Doi: 10.60627/0P45-ZJ55*. (2023).

# Switching Graphitic Polytypes in Elastically Coupled Islands

# Supplementary Information

### SI.1. Device Fabrication Details

Our device structure consists of (from top to bottom); a bi-layer graphene (BLG) / misaligned monolayer graphene (MLG) spacer with a cavity array / aligned BLG / 50 nm thick hBN/ 90nm SiO$_2$ / 0.5 mm Si substrate, fabricated using a dry polymer stamping method[1]. A large BLG is initially etched into two parts using high-intensity pulsed laser light (1064 nm, WITEC alpha300 Apyron, Fig.S1.a, d). In a separate step, we etch a cavity array into a MLG spacer using electrode-free local anodic oxidation (LAO) lithography[2] (Fig.S1.b, e).

We assembled the device using a Polyvinyl alcohol (PVA)-coated polydimethylsiloxane (PDMS) stamp to pick up the hBN, the first BLG, the MLG spacer, and the second BLG. To flip the structure and place the active BLG at the top, facing the tip of the AFM, we use a second PDMS stamp coated with PMMA (Polymethyl methacrylate). Then we bring the stacked layers from the PDMS-PVA stamp into contact with the PMMA stamp and inject water into the gap to dissolve the PVA coating, facilitating the transfer[3]. Finally, we drop the entire stack onto a Si/SiO$_2$ substrate at 150°C. After fabrication, we design electrical contact pads (Cr/Au) using electron beam lithography.

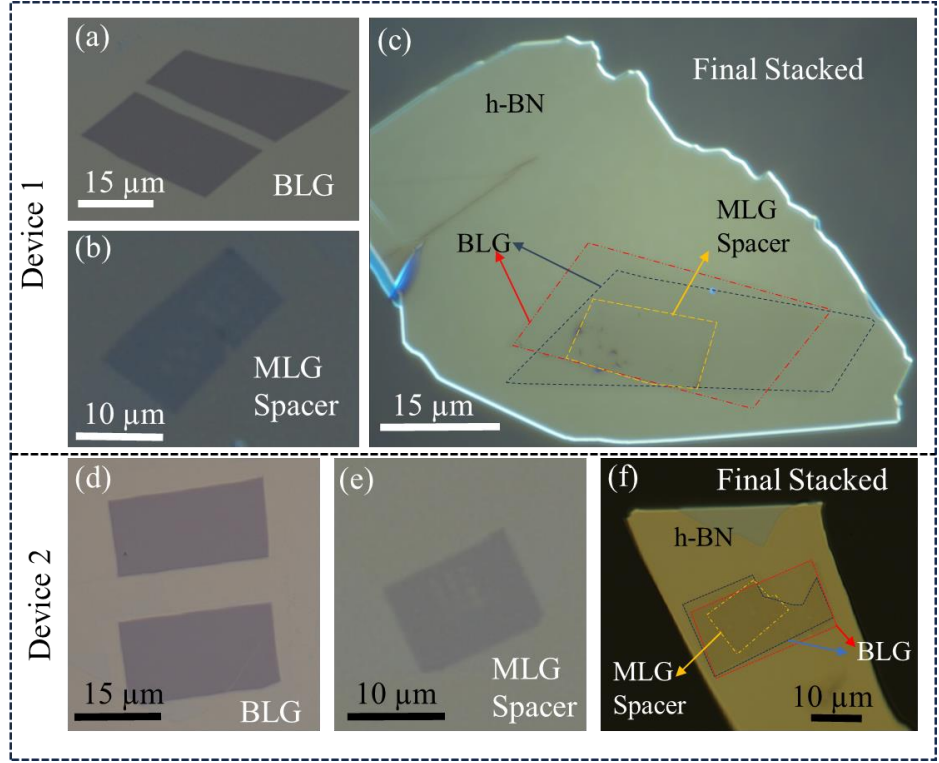

*Figure S1: Optical image of the devices. (a) Two pieces of bilayer graphene (BLG), (b) Monolayer graphene spacer (MLG), (c) Final device, where the red, blue, and yellow dashed rectangles represent the boundaries of the two BLG pieces and the MLG spacer, respectively. (d-f) Another similar device.*

## SI.2. Conducting Atomic Force Microscopy (C-AFM) Measurements

The current maps presented in this work were captured using a Park NX10 Hivac in C-AFM mode under an $N_2$ atmosphere. A HQ:NSC35/Pt tip with a spring constant of 5.4 N/m and a resonant frequency of ~150 kHz was used for all measurements. Prior to current mapping, the sample surface was thoroughly cleaned by performing multiple scans in contact mode with a loading force of 350 nN. This process removes any residual contaminations and confirms that all cavity arrays are properly sagged. Following the tip cleaning, a fresh AFM tip is utilized for the current imaging. To avoid saturation of the tip current due to low resistance in several specific devices, we add a series resistor (1-10 MΩ) into the measurement circuit.

## SI.3. Friction Force Imaging

We record the lateral photodiode signals during both forward and backward scans, as illustrated in Fig.S2.a, b, to measure the friction force. Friction force induces torsional motion in the AFM tip (Fig.S2.b), which manifests as a difference between the forward and backward lateral photodiode signals (measured in volts), as shown in Fig.S2.c and d. We generate a friction force map by calculating this difference, which reflects the torsional motion of the AFM tip. To convert the potential shift into a friction force, we use a $SiO_2$ substrate (with a known friction coefficient of 0.08) and a known conversion factor (nN/V) for lateral force, see Ref[4]. Fig.S2.e shows a plot of the friction force as a function of the loading force, as we deduce from dedicated friction force maps (see examples in panels f-h).

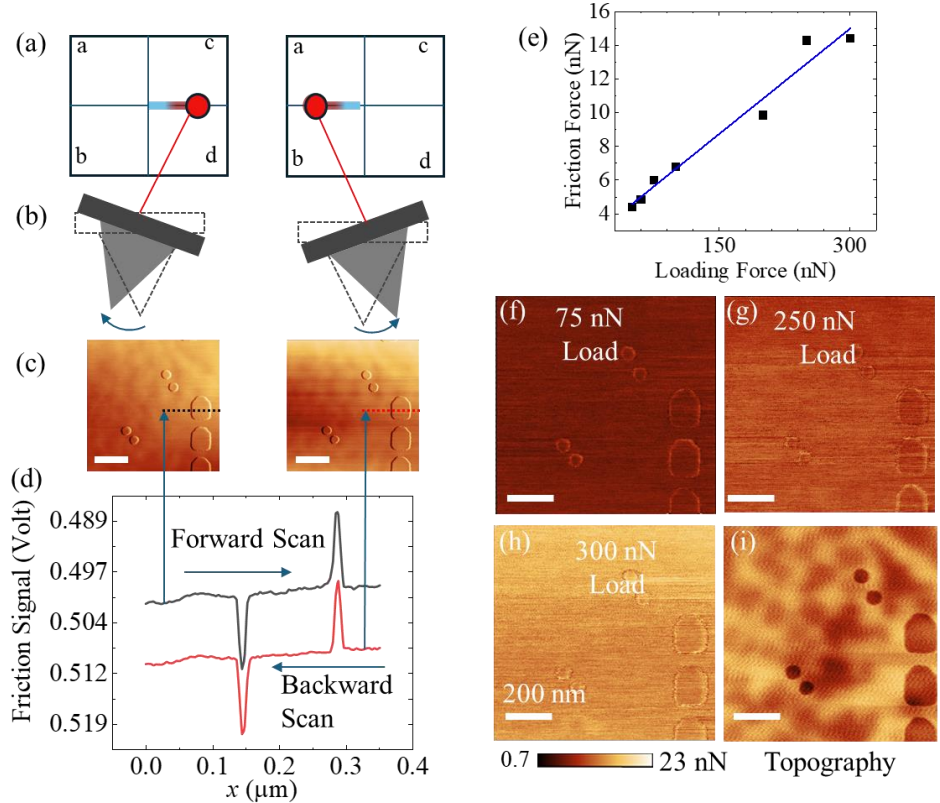

*Figure S2: **Friction force imaging.** (a, b) Schematic illustration of lateral photodiode signals and the torsional motion of the AFM tip during forward and backward scans. (c) Map of the lateral signal for forward and backward scans. (d) Line-cut along the dashed line in (c). (e) Friction force as a function of the loading force. (f–h) Examples of friction force calibration maps. (i) Corresponding topography map.*

### SI.4. Identifying Bernal (B) and Rhombohedral (R) Polytypes

To confirm the B / R stacking order within each cavity, we conducted 2D Raman mapping and current mapping on naturally exfoliated tetra-layer graphene samples. Fig.S3.a shows an optical image example of a typical flake. Panel (b) shows a Raman map of the red-framed area, which is constructed by measuring the reflection spectrum at each pixel and assigning intensities based on the integrated light intensity at the filtered spectrum range shown by the gray bar in panel c. The spectra taken within the domains (using a green laser), confirm the presence of B and R polytypes in the measured regions[5].

The corresponding current map of the same regions is displayed in Fig.S3.d. R regions exhibit higher current compared to the B regions as in the case of the SLAP devices (see panels e, f). We note that the R domains tend to shrink during the cleaning process using contact mode scanning. We observed the same Raman / current correspondence in five different samples.

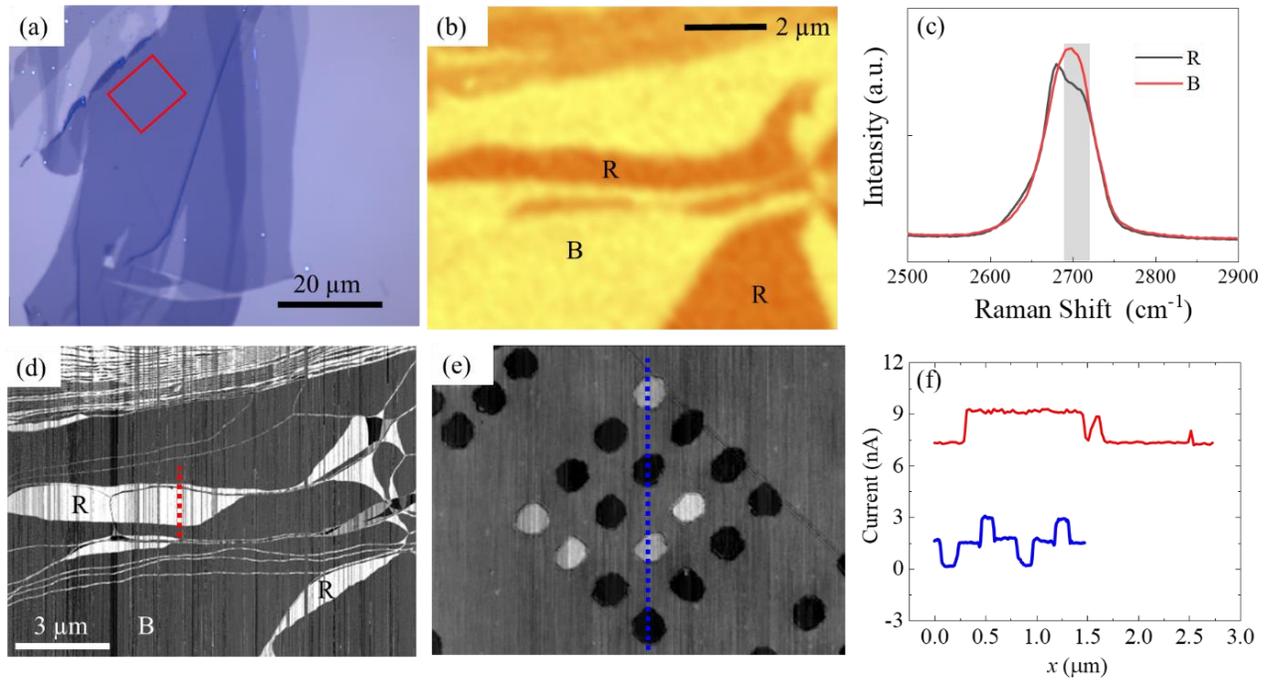

*Figure S3: Comparison of Raman and current maps.* *(a) Optical image of few-layer exfoliated graphene flakes. (b) 2D Raman map of the four-layer graphene region outlined by the red frame in (a). (c) 2D Raman peaks corresponding to R and B polytype regions. (d) Current map of the regions shown in (b). (e) Current map of the SLAP device. (f) Line-cut along the dashed lines in (d) and (e).*

**SI.5. Effect of Spacer Commensuration on the Switching Dynamics**

To test the influence of the lubricant medium between the islands on the structural switching dynamics within the cavities, we studied two devices with aligned MLG spacer. Typically, twist angles of more than 5 degrees show well-developed structural lubricity while marginal twists of less than 0.3 degrees show triangular patterns of commensurate domains with moiré wavelength $\lambda > 50$ nm. In the case of intermediate twists, one expects partial commensuration and planar relaxations into hexagonal moiré patterns rather than triangular domains[6].

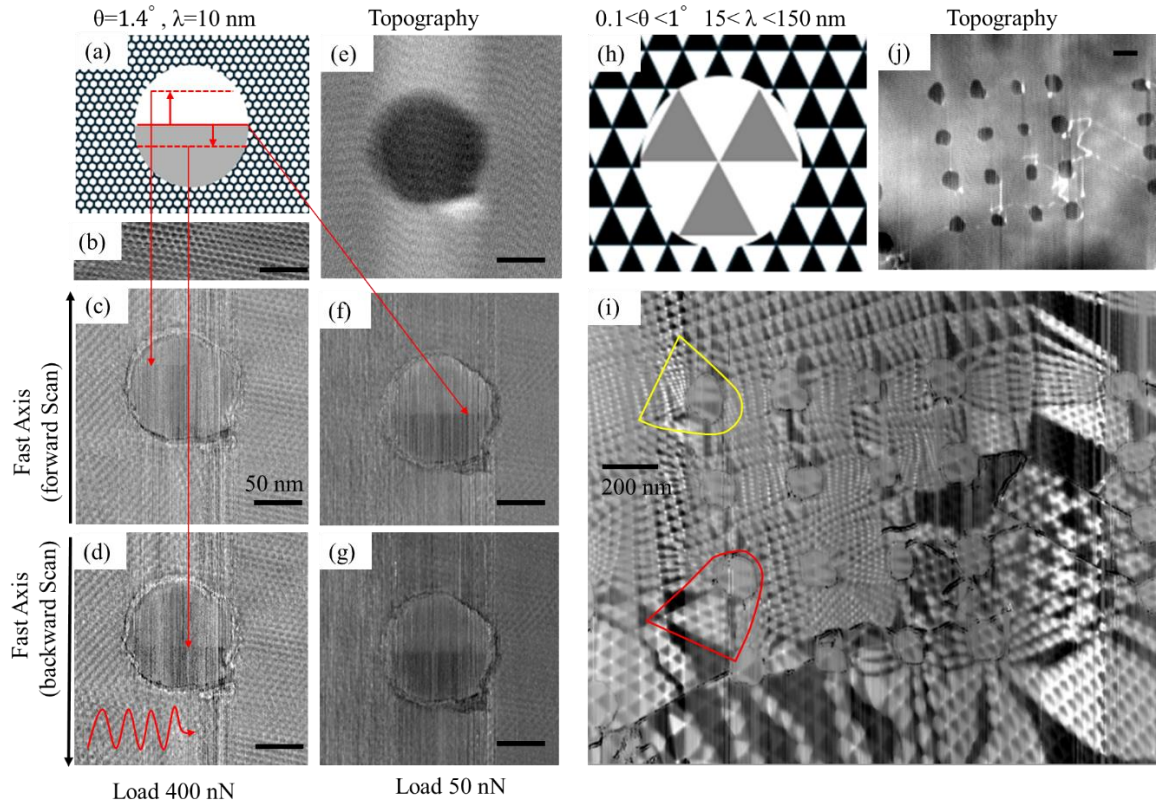

*Figure S4: Strip dynamics versus spacer twist angle.* (a) Illustration of a dense dislocation pattern with short, partially commensurate, hexagonal moiré wavelength λ=10 nm between the spacer and the aligned active bilayers. (b) Zoom in current map near the cavity. (c, d) Current maps collected at forward (c) and backward (d) scan direction under a vertical tip load of 400 nN. In each successive line, the B/R boundary strip in the cavity is shifted before it retracts back spontaneously to its initial middle position. (e) Topography of the measured area. (f, g) Similar current maps using low 50 nN load, with friction force below the coercive force value. (h) Illustration of a commensurate and hence triangular moiré pattern with the spacer. (i) Current map of a commensurate cavity array. Although the twist between the active bilayers is smaller than with the spacer, it remains finite inside the cavity (see islands with triangles). The red frame outside the cavity marks a large triangle (with no twist between the active layers) containing small triangles due to the twist with the spacer. Here, the commensurate patterns outside the holes stabilize moiré patterns within the holes. The yellow frame shows smaller domains outside the cavity with a less commensurate spacer. (j) Topographic image of the area in (i).

Fig.S4 shows examples of partially and highly commensurate spacers. The λ= 10 nm moiré wavelength in panels a-g corresponds to a twist angle of θ=1.4° between the spacer and the active bilayer outside the cavity. In this intermediate twist range with reduced lubricity, we typically find a few R to B boundaries that remain stable inside the cavity (rather than a spontaneous strip motion to the cavity edge). Hence, the residual pinning outside the cavities suppresses the structural switching within the cavity. More specifically, in response to large loading and friction forces, the boundary strips move only temporarily during scanning and revert to their original positions once the tip is lifted (see the two dashed red lines in panel a). Panels c and d show forward scan and backward scan maps respectively (under 400 nN load) in which the strip moves forward and backward multiple times in each successive line scan. Panels f and g show the same forward/backward scans with a lower load (~50 nN), where the dislocation remains pinned to the same position.

Panels h,i show another device with a more commensurate spacer and larger moiré wavelength $15 < \lambda < 150$ nm, corresponding to marginal twist angles in the range $0.1° < \theta < 1°$ at different sections of the sample. In addition, the active layers are twisted by $0.01° < \theta_{active} < 0.2°$. These two twisted interfaces result in complicated domain patterns of penta-layer polytypes between the islands. Nevertheless, it is possible to identify the relative structural stability in different sections of this device by comparing the domain dimensions inside and outside the cavities. First, we point to the red frame of panel i. The triangles in this cavity show a $\lambda \sim 100$ nm length ($\theta_{active} \sim 0.14°$), while outside the cavity, the active layers create a relatively large triangle with $\lambda \sim 250$ nm ($\theta_{active} \sim 0.06°$) which is divided into smaller triangles with $\lambda \sim 50$ nm ($\theta \sim 0.3°$) by the twisted spacer. Now compare this with the yellow frame. Here the cavity also shows a $\lambda \sim 100$ nm length ($\theta_{active} \sim 0.14°$), but the active layers outside the cavity create relatively smaller triangles with $\lambda \sim 50$ nm ($\theta_{active} \sim 0.3°$) which are divided into tiny domains with $\lambda \sim 15$ nm ($\theta \sim 1°$) by the more twisted spacer. It means that the boundary strips between the active bilayers tend to move out of the top cavity due to the higher lubricity of the spacer. Conversely, in the bottom red frame, they tend to slide into the cavity which is here embedded in a commensurate tetralayer medium. This behavior highlights the significant impact of the misaligned spacer on the relative domain stability and mechanical switching of the SLAP devices. Aligned spacers cause strong pinning that limits the strain relaxation length, reduces the dislocation motion, and freezes the switch dynamics.

**References:**

1. Onodera, M. *et al.* All-dry flip-over stacking of van der Waals junctions of 2D materials using polyvinyl chloride. *Sci Rep* **12**, (2022).

2. Li, H. *et al.* Electrode-Free Anodic Oxidation Nanolithography of Low-Dimensional Materials. *Nano Lett* **18**, 8011–8015 (2018).

3. Li, Y. *et al.* Symmetry Breaking and Anomalous Conductivity in a Double-Moire Superlattice. *Nano Lett* **22**, 6215–6222 (2022).

4. Bosse, J. L., Lee, S., Andersen, A. S., Sutherland, D. S. & Huey, B. D. High speed friction microscopy and nanoscale friction coefficient mapping. *Meas Sci Technol* **25**, (2014).

5. Atri, S. S. *et al.* Spontaneous Electric Polarization in Graphene Polytypes. *Advanced Physics Research* **3**, (2024).

6. Vizner Stern, M., Salleh Atri, S. & Ben Shalom, M. Sliding van der Waals polytypes. *Nature Reviews Physics* (2024) doi:10.1038/s42254-024-00781-6.